\documentclass[12pt,twoside,pointlessnumbers,smallheadings]{revtex4}
\usepackage{amsmath}
\usepackage{amssymb}
\usepackage{color}

\usepackage{hangcaption,fancybox,feynmp}
\usepackage{indentfirst}
\usepackage{graphicx}
\usepackage{epsfig,subfigure,psfrag}
\usepackage{CJK}
\usepackage{mathrsfs}

\begin{document}


\title{
$\mathscr{O}(100 GeV)$ Deci-weak  $W^\prime/Z^\prime$ at Tevatron and LHC
}
\author{Xiao-Ping Wang $^{1}$, You-Kai Wang $^{1}$, Bo
Xiao$^{1}$, Jia Xu $^{1}$ and Shou-hua
Zhu$^{1,2}$}

\affiliation{ $ ^1$ Institute of Theoretical Physics $\&$ State Key
Laboratory of Nuclear Physics and Technology, Peking University,
Beijing 100871, China \\
$ ^2$ Center for High Energy Physics, Peking University, Beijing
100871, China }

\date{\today}

\maketitle

\begin{center}

{\bf Abstract}

\begin{minipage}{15cm}
{\small  \hskip 0.25cm

Recently Tevatron released their measurements on invariant mass spectrum of electron/positron, as well as the di-jet arising from
WW+WZ production with one W leptonically decay. Though the statistics is not significant, there are two bumps around 240 GeV and 120-160 GeV
respectively. We proposed that the two bumps correspond to the extra light gauge bosons $Z^\prime$ and $W^\prime$, which couple with
quarks with the deci-weak strength. In this brief report, we also simulated di-jet invariant mass distribution at the current running LHC.

}

\end{minipage}
\end{center}


\newpage

\section{Introduction\label{introduction}}

Extra gauge bosons $W^\prime$ and $Z^\prime$ are theoretical well motivated. For example, the left-right symmetric models \cite{Mohapatra:1974hk,Senjanovic:1975rk} or the little
Higgs model \cite{ArkaniHamed:2002qy} contain the extra gauge group $SU(2)$. Even before the electro-weak symmetry breaking (EWSB),
the extra gauge bosons obtain mass via some unknown mechanism. Usually there are also the addition contributions from the EWSB.
The null signal from observation will push the scale of unknown mechanism to O(TeV). Consequently the  $W^\prime$ and $Z^\prime$ are at O(TeV). However Tevatron
recent measurements may point to the alternative way, namely $O(100 GeV)$ $W^\prime$ and $Z^\prime$ with couplings with SM particles weaker than weak interaction.

CDF/D0 at Tevatron released recently their searches on new resonances which decay into electron and positron \cite{Aaltonen:2008vx,Abazov:2010ti}.
There seems an unexpected bump around 240 GeV in the di-electron invariant
spectrum, though the significance is only $3.8\sigma$ \cite{Chanowitz:2011ew}. The author of Ref. \cite{Chanowitz:2011ew} attributed the bump to the extra $Z^\prime$ in
the little Higgs models. He found that the low mass $Z^\prime$ is consistent with other precise electro-weak data, though the coupling between $Z^\prime$ and
SM fermions are highly suppressed (fine-tuned?). However the best-fit prefers the heavy Higgs boson which contradicts with the origin motivation to construct
the little Higgs model.

We noticed that CDF/D0 at Tevatron released recently also their measurements on WW/WZ cross sections via the
electron/muon + missing energy + di-jet channel \cite{Aaltonen:2009vh,CDFemu2jetlatest,cdflatest}.
The cross sections are in agreement with the standard model (SM) prediction. However there seems an unexpected bump in the di-jet
spectrum around 120-160 GeV, though the significance ($3.2 \sigma$ \cite{cdflatest})is not significant. Is that possible that such bump arises from
the extra $W^\prime$, at the same time the 240 GeV bump is due to the extra $Z^\prime$?
In order to escape the severe constraints from past measurements, the couplings among extra light gauge bosons with SM fermions usually highly suppressed, as
the case of $Z^\prime$ \cite{Chanowitz:2011ew} in the little Higgs model.
However the couplings among $Z^\prime - W^\prime - W$ are not necessary severely suppressed. Once such decay channel is opened, $Z^\prime$ branching ratio to electron/positron
can be suppressed. As such, the coupling among $Z^\prime$ and SM fermions, especially quarks, might not so severely suppressed as that in the little Higgs model \cite{Chanowitz:2011ew}. The coupling strength
may shift from centi-weak  \cite{Chanowitz:2011ew} to deci-weak.

In this brief note, we will treat two $\mathscr{O}(100 GeV)$  bumps as the signals of extra $Z^\prime$  and  $W^\prime$ via
\begin{eqnarray}
p\bar p \rightarrow Z^\prime \rightarrow e^+e^- {\it \ or \ } W^\prime +W \end{eqnarray}
respectively. The further cross-check at LHC is also scrutinized.

\section{Extra $W^\prime$ and $Z^\prime$ at Tevatron}

As shown in Ref. \cite{Chanowitz:2011ew}, the cross section for $e^+e^-$ bump production is
\begin{eqnarray}
\sigma (p\bar p\rightarrow Z^\prime) Br(Z^\prime \rightarrow e^+e^-) \simeq 42\ fb
\end{eqnarray}
with $m_{Z^\prime} =240 GeV$. Based on the subtraction figure in \cite{cdflatest}, the extra contribution can be read
roughly
\begin{eqnarray}
\sigma (p\bar p\rightarrow Z^\prime) Br(Z^\prime \rightarrow W W^\prime) Br( W^\prime \rightarrow jj)  \sim \mathscr{O} (1~pb)
\end{eqnarray}
with $m_{W^\prime} =120-160 GeV$.
In order to fill the gap between $e^+e^-$ and di-jet productions, we require
\begin{eqnarray}
\frac{Br(Z^\prime \rightarrow W W^\prime) Br( W^\prime \rightarrow jj)}{ Br(Z^\prime \rightarrow e^+e^-)} \gg 1.
\label{ratio}
\end{eqnarray}

In principle, we don't know yet the structure and coupling strength of SM fermions with $W^\prime$/$Z^\prime$, as well as $W-Z^\prime-W^\prime$. From Eq. \ref{ratio}, it is natural to require that couplings of $Z^\prime$ and leptons are less than those with quarks and gauge bosons. In fact the couplings among leptons and extra gauge bosons are severely constrained due to the precise measurements by LEP and
other experiments. Therefore we focus on the couplings among quarks and extra gauge bosons. For simplicity,
we choose the coupling structure the same as that of SM, but with an additional suppression factor $\kappa_{Z^\prime f \bar f}$, $\kappa_{W^\prime q \bar q}$ and
$\kappa_{W-Z^\prime-W^\prime}$ respectively.

\begin{figure}[htbp]

\begin{center}
\includegraphics[width=0.80\textwidth]
{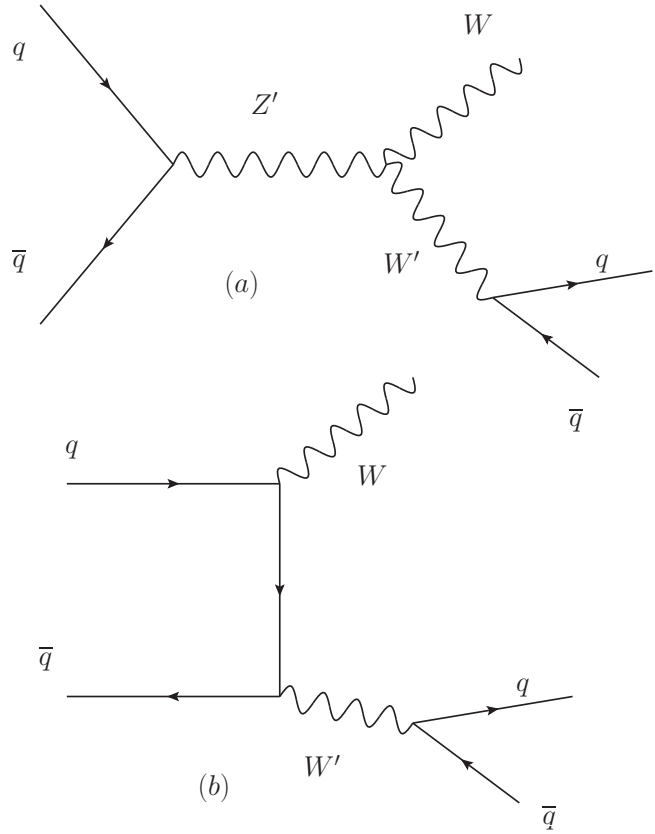}
\end{center}

\caption{\label{fdww} Feynman diagrams for $W^\prime$ production at Tevatron and LHC.
}

\end{figure}

In order to account for the CDF data, we simulate the di-jet invariant distribution arising from $WW/WZ$ as well as $W^\prime W$ production with $m_{Z^\prime}=240$ GeV and $m_{W^\prime}=150$ GeV respectively. For the latter case there are two Feynman diagrams at tree level as depicted in Fig. \ref{fdww}.

\begin{figure}[htbp]

\begin{center}
\includegraphics[width=0.80\textwidth]
{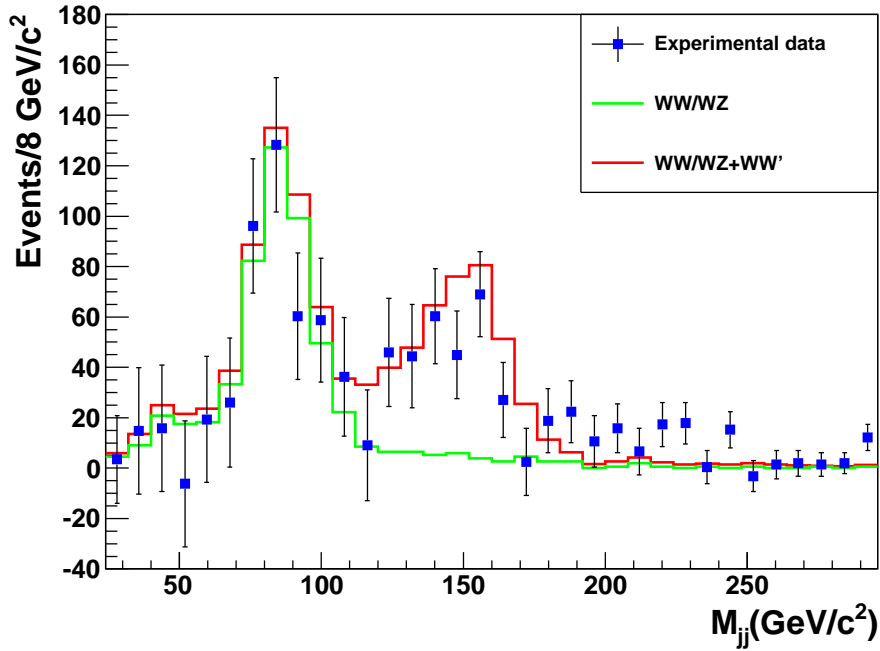}
\end{center}

\caption{\label{mjjatcdf} Di-jet invariant mass distribution at Tevatron with integrated luminosity $\mathscr{L}=4.3~fb^{-1}$. Data is taken from Ref. \cite{cdflatest}.
}

\end{figure}

We choose the benchmark parameter set as $\kappa_{Z^\prime f \bar f}=0.07$, $\kappa_{W^\prime q \bar q}=0.3$ and
$\kappa_{W-Z^\prime-W^\prime}=0.1$, which do not contradict with other measurements \cite{Chanowitz:2011ew,Langacker:1989xa,Grojean:2011vu}. The di-jet invariant distribution is shown
in Fig. \ref{mjjatcdf} after imposing the same cuts with those of \cite{cdflatest}. The events are generated
by MadGraph \cite{Maltoni:2002qb}, then the initial state radiation, final state radiation and fragmentation are carried out by Pythia \cite{Sjostrand:2006za}. The detector response is simulated by PGS. We corrected the jet energy according to Ref. \cite{phdthesis}.
From the Fig. \ref{mjjatcdf}, we can see that the new contribution arising from
the extra $W^\prime$ can excellently fit the data.  We should emphasize that the di-jet $P_T$ cut has suppressed
the contributions from s-channel diagram (a) in Fig. \ref{fdww}. The contributions to the new bump comes mainly
from the t-channel diagram (b).

\section{ Extra $W^\prime$ and $Z^\prime$ at LHC}

Since Tevatron will be closed soon. The further study on such new gauge bosons at the LHC is the important task.
The $Z^\prime $ can be copiously produced at LHC. For the benchmark parameter $\kappa_{Z^\prime q \bar q}=0.07$,
the cross section for $Z^\prime$ is around 6 pb for $\sqrt{s}=7$ TeV. In order to induce the observed $e^+e^-$ events at Tevatron, the $Br (Z^\prime \rightarrow e^+e^-) \simeq 2.7\%$. Thus the effective cross section for $e^+e^-$ production is around 180 fb at the LHC.

\begin{figure}[htbp]

\begin{center}
\includegraphics[width=0.80\textwidth]
{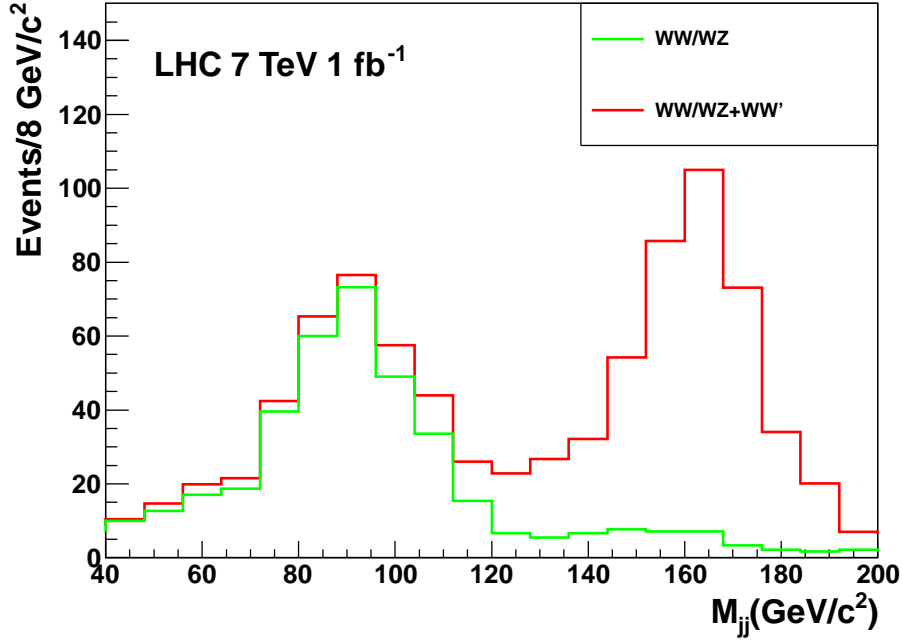}
\end{center}

\caption{\label{mjjatlhc} Di-jet invariant mass distribution at LHC with $\sqrt{s}=7$ TeV and the integrated luminosity $\mathscr{L}=1~fb^{-1}$.
}

\end{figure}

Similar to the case at Tevatron, $W^\prime$ will appear as the bump of di-jet associated with isolated electron/muon plus missing energy
\begin{eqnarray}
p p \rightarrow Z^\prime \rightarrow W(\rightarrow e/\mu + \nu)+ W^\prime (\rightarrow jj)
\end{eqnarray}
In order to illustrate the signal at LHC, in Fig. \ref{mjjatlhc} we plot the di-jet invariant mass distribution with $\sqrt{s}=7$ TeV. The selection rules are the same as those of Tevatron. From the figure it is obvious that signal events from $W^\prime$ have the similar magnitude with those of SM. Such extra bump should be very clearly identified for the early LHC data.

\section{Conclusions and discussions }

In this brief report, we treat the two bumps observed by Tevatron, as the possible $W^\prime$ and $Z^\prime$ signal. Based on our investigations,
it is not hard to discover/exclude such extra gauge bosons at the LHC, even their coupling with SM fermions are suppressed. It is natural to quest the possible
origin of such extra gauge bosons. Obviously the traditional left-right symmetric model can't be the candidate because of the large couplings with
the SM fermions. The little Higgs model is not preferred, though in some case the 240 GeV bump can be accounted for. Recently man expect that
there is complicated 'hidden world', which contains extra non-abelian gauge group. After symmetry breaking, the gauge boson eigenstates are
the mixed states from both gauge fields. The construction of such kind of model is emergent once the
signal observed at Tevatron is confirmed.
Last but not least even the two bumps are real signal, they may arise from new particles other than gauge bosons. In order to pin down the spin of new particle,
one need collect more data to analyze the angular distribution of final states it decays into.

{\em Note added:} While this paper is in writing, we aware that there are several papers \cite{Bai:2010dj,Buckley:2011vc,Yu:2011cw}  before CDF released their measurement 
\cite{cdflatest}.

\section*{Acknowledgment}

We thank Jia Liu for drawing our attention to the measurements of WW/WZ cross section by
Tevatron. This work was supported in part by the Natural Sciences Foundation
of China (No 11075003).


\end{document}